\documentclass[3p,12pt]{elsarticle}

\usepackage{verbatim}

\usepackage[colorlinks=true]{hyperref}
\hypersetup{urlcolor=blue, citecolor=red}

\usepackage{graphicx}
\usepackage{epstopdf}
\usepackage{floatrow}
\usepackage{subfig}
\usepackage{multirow}
\usepackage{multicol}

 \usepackage{amsmath,amsfonts,amssymb,amsthm,latexsym}
 
 \biboptions{numbers,sort&compress}

\newcounter{mnote}
\theoremstyle{plain}

\theoremstyle{definition}

\theoremstyle{remark}

\makeatletter
\def\ps@pprintTitle{%
 \let\@oddhead\@empty
 \let\@evenhead\@empty
 \def\@oddfoot{\centerline{\thepage}}%
 \let\@evenfoot\@oddfoot}
\makeatother

\def\CO2{CO$_2$}

\begin{document}

\begin{frontmatter}

\title{Rayleigh fractionation in high-Rayleigh-number solutal convection in porous media}

\author{Baole Wen}
\ead{wenbaole@gmail.com or baole@ices.utexas.edu}
\author{Marc A. Hesse}
\ead{mhesse@jsg.utexas.edu}
\address[ICES]{Institute of Computational Engineering and Sciences, The University of Texas at Austin, Austin, TX 78712 USA}
\address[GEO]{Department of Geological Sciences, Jackson School of Geosciences, The University of Texas at Austin, Austin, TX 78712 USA}


\begin{abstract}
We study the fractionation of two components between a well-mixed gas and a saturated convecting porous layer. Motivated by geological carbon dioxide (\CO2) storage we assume that convection is driven only by the dissolved concentration of the first component, while the second acts as a tracer with increased diffusivity. Direct numerical simulations for convection at high Rayleigh numbers reveal that the partitioning of the components, in general, does not follow a Rayleigh fractionation trend, as commonly assumed. Initially, increases in tracer diffusivity also increase its flux, because the diffusive boundary layer penetrates deeper into the flow. However, for $D_2\geq 10\, D_1$, where $D_1$ and $D_2$ are, respectively, the diffusion coefficients of \CO2 and the tracer in water, the transverse leakage of tracer between up- and down-welling plumes reduces the tracer flux. Rayleigh fractionation between components is only realized in the limit of two gases with very large differences in solubility and initial concentration in the gas. 
\end{abstract}



\end{frontmatter}

\section{Introduction}
\label{sec:intro}
Convection in porous media controls many mass and heat transport processes in nature and industry \cite{NB2006} and Rayleigh-Darcy convection is also a classic example of spatiotemporal pattern formation \cite{Rayleigh1916,Cross1993}. This subject has received renewed interest due to its potential impact on geological carbon dioxide (\CO2) storage. The injection of supercritical \CO2 into deep saline aquifers for long-term storage is the only technology that allows large reductions of \CO2 emissions from fossil fuel-based electricity generation \cite{Orr2009,Michael2010,Szulczewski:2012,Liang2017b,Liang2017a}. Dissolution of \CO2 into the brine eliminates the risk of upward leakage \cite{Gasda:2004,Roberts2011,Trautz2013}, because it increases the density of the brine and forms a stable stratification \cite{Weir:1996}.

Once the diffusive boundary layer of dissolved \CO2 in the brine has grown thick enough it becomes unstable and convective mass transfer allows a constant dissolution rate \cite{Ennis-KingPaterson:2005,Riaz2006,Neufeld2010}. The time scale for the onset in typical storage formations is at most a few centuries \cite{Ennis-King2005,Riaz2006,Wessel-Berg:2009,Slim2010}, so that convective mass transport determines the rate of \CO2 dissolution. Recent work has therefore focused on determining the convective dissolution rate in numerical simulations \cite{Hassanzadeh:2007,Pau2010, Hidalgo2012, Hewitt2012, Slim2013, Fu:2013, Hewitt2013shutdown, Wen2012, Wen2013, Shi2017, Wen2018JFM} and laboratory experiments \cite{Neufeld2010,Backhaus:2011,Tsai:2013,Liang2017thesis}.


However, most of these studies consider convection in homogeneous porous media, while geological formations exhibit extreme heterogeneity at all scales \cite{Sposito2008,Yeh2015}. It is therefore important to complement numerical and experimental work with estimates of convective dissolution rates in real media that have been inferred from field observations. All such estimates are based on increases in the abundance of Helium (He) relative to \CO2 in the residual gas, as convection strips the more soluble \CO2 \citep{Cassidy2005, Gilfillan2008, Gilfillan2009, Sathaye2014, Sathaye2016a, Sathaye2016b}. These studies interpret the observed changes in the \CO2/He ratio in terms of a zero-dimensional Rayleigh fractionation model \cite{Rayleigh1896,Rayleigh1902,White2013,Clark2015}. 

This interpretation assumes that the fractionation depends only on the solubility of the components, but not on their diffusion coefficients. In the absence of convection, however, mass transfer is controlled by diffusion and this assumption must break down. In a strongly convecting fluid, in contrast, advective mass transfer is dominant and differences in diffusivity may become negligible. One might therefore expect Rayleigh fractionation between solutes in the limit of high-Rayleigh-number convection. Here, we directly test this hypothesis using highly resolved direct numerical simulations (DNS) of solutal convection in a porous medium. However, unlike the double-diffusive (or combined thermal and solutal) convection \cite{NB2006}, the convection considered here is only driven by the buoyancy force due to the density change induced by the first solute (\CO2). Despite the simplicity of this physical system the emergence of complex behavior is observed.

The manuscript is structured as follows. First, we obtain an expression for the evolution of the residual gas composition as a function of the convective fluxes of the two components in the liquid. These fluxes are then obtained from DNS of high-Rayleigh-number solutal convection in a porous medium. Finally, we determine the conditions under which the residual gas composition experiences Rayleigh fractionation.



\section{Problem formation and computational methodology}
\label{sec:Problemformation}
\begin{figure}[t!]
    \centering
    \includegraphics[width=0.65\textwidth]{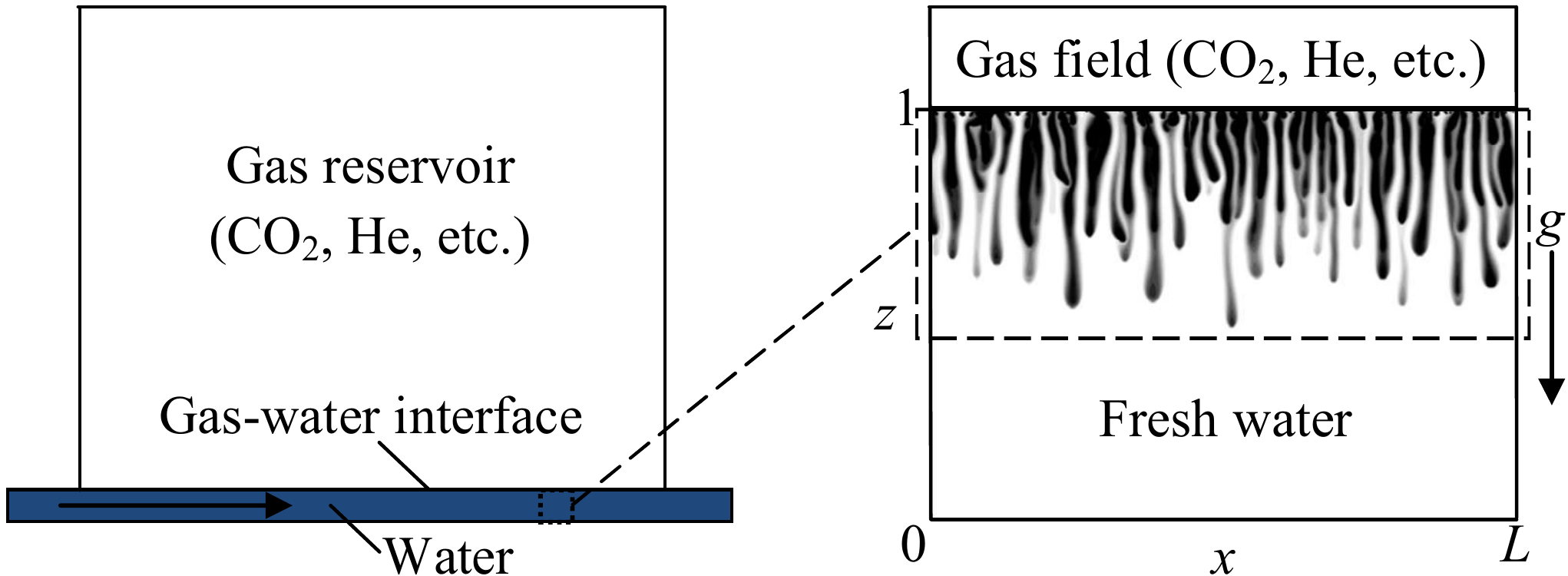} 
  \caption{Schematics showing the Rayleigh fractionation process in simple geometries.  The assumed physical mechanism leading to Rayleigh fractionation is convection (advection), because it continuously brings in new brine that is saturated at the gas-water interface and subsequently removed.}  \label{Schematics}
\end{figure}

In a binary system, the composition of the gas is characterized by the ratio of moles between CO$_2$ and the tracer (i.e. He) in the gas field, $r = n_{1,g}/n_{2,g}$, where the subscripts `1' and `2' denote the solutes \CO2 and He, respectively, and `$g$' the gas phase. This gas is in contact with a convecting fluid that equilibrates instantaneously at the gas-water interface and constantly removes the dissolved components and carries new unsaturated water to the interface (Fig.~\ref{Schematics}). The change of the $i$-th component ($i = 1$, 2 here) in the gas is therefore given by 
\begin{eqnarray}
 \dfrac{dn_{i,g}}{dt} = -F_i\frac{D_1^*C_{is}}{H} A, \label{ODE_Rayleigh}
\end{eqnarray}
where $F_i$ is the corresponding dimensionless flux defined later in Eq.~(\ref{Flux}), $D_1^*$ is the dimensional diffusivity for the first solute, $C_{is}$ is the saturated concentration of the $i$-th component in the water, $H$ is the thickness of the water layer and $A$ is the gas-water contact area. We assume an open system  in contact with a liquid reservoir at constant pressure. This implies that the pressure in the gas remains constant as dissolution proceeds, but the gas volume declines. Further we assume that the gas is ideal and that partitioning is described by Henry's law \cite{Henry1803c}. High Rayleigh-number convection is quasi-stationary so that the convective flux $F_i$ is constant. Following \cite{White2013} and \cite{Criss1999} the fraction of the initial \CO2 that has dissolved into the water is given by
\begin{eqnarray}
 \mathcal{F} \equiv 1 - {n_{1,g}}/{n_{1,g}^0} = 1 - (r/r^0)^{\frac{\alpha}{\alpha - 1}}, \label{Fraction_Rayleigh}
\end{eqnarray}
where the superscript `$0$' denotes the initial state. The evolution of the gas composition is governed by the fractionation factor, 
\begin{align}
\alpha = \frac{F_1K_1}{F_2K_2}, \label{alpha}
\end{align}
where $K_i$ is Henry's law solubility constant of the $i$-th component (see the detailed derivation in the Appendix section).  In the limit of Rayleigh fractionation the fluxes for different solutes are assumed to be identical, $F_{1} \equiv F_{2}$, so the Eq.~(\ref{alpha}) becomes $\alpha = K_1/K_2$.

To determine these convective fluxes we study the Boussinesq, Darcy flow in a dimensionless 2D porous layer with horizontal and vertical coordinates $x$ and $z$, respectively, as shown in Fig.~\ref{Schematics}.  We assume the density-driven flow $\mathbf{u} = (u, w)$ through the homogeneous and isotropic porous media is incompressible \cite{NB2006},
\begin{eqnarray}
& \mathbf{u} = -{\nabla}p - \text{Ra}(C_1 + {\beta}C_2){\bf e}_{z}, \label{Momentum_nondim} \\
&\nabla\cdot\mathbf{u} = 0, \label{Continuity_nondim}\\
& \dfrac{\partial C_i}{\partial t} + \mathbf{u}\cdot\nabla C_i = D_i{\nabla}^2 C_i, \quad i = 1, 2, \label{Solute_nondim}
\end{eqnarray}
where ${p}$ is the pressure field, ${\bf e}_z$ is a unit vector in the $z$ direction, $C_i$ and $D_i$ are, respectively, the concentration and diffusivity of the $i$-th solute, $\beta$ is the weighting factor of buoyancy force for $C_2$, and the Rayleigh number $\mbox{Ra} = HK\text{g}\triangle\rho_1/(\mu{\varphi}D_1^*)$ where $K$ is the medium permeability, $\text{g}$ is the acceleration of gravity, $\triangle\rho_1$ is the density difference between the fresh water and the saturated water for the first solute, $\mu$ is the dynamic viscosity of the fluid, and $\varphi$ is the porosity.  Since $D_1^*$ is used for normalization of time, $D_1 \equiv D_1^*/D_1^* = 1$, and $D_2 \equiv D_2^*/D_1^*$ is the ratio of diffusivities between the two solutes.  
Here, the second solute $C_2$ is a passive tracer which does not change the density of the brine, so $\beta = 0$.  For boundary conditions, the lower boundary is impenetrable to the fluid and solutes, the upper boundary is saturated (i.e., $C_i = 1$) and impenetrable to the fluid, and all fields are $L$-periodic in $x$.  One of the key quantities of interest in solutal convection is the dissolution flux $F$ representing the rate at which the solutes dissolve from the upper boundary of the layer, defined as 
\begin{eqnarray}
F_{i}(t) =  \dfrac{D_i}{L}\int^{L}_0\left.\dfrac{\partial C_i}{\partial z}\right\vert_{z=1}dx \;\; \mbox{for} \;\; i = 1, 2, \label{Flux}
\end{eqnarray}
where $L$ is the aspect ratio of the domain.  

The equations~(\ref{Momentum_nondim})--(\ref{Solute_nondim}) are solved numerically using a Fourier--Chebyshev-tau pseudospectral algorithm \citep{Boyd2000}.  For temporal discretization, a third-order-accurate semi-implicit Runge--Kutta scheme \citep{Nikitin2006} is utilized for computations of the first three steps, and then a four-step fourth-order-accurate semi-implicit Adams--Bashforth/Backward--Differentiation scheme \citep{Peyret2002} is used for computation of the remaining steps, so generally it is fourth-order-accurate in time.  We performed computations for a discrete set of Rayleigh number and ratio of diffusivities from Ra $= 50$ to Ra $= 5\times10^4$ and $D_2 = 1.25$ to $D_2 = 100$ in the 2D domain with aspect ratio $L = 10^5/\mbox{Ra}$.  8192 Fourier modes were utilized in the lateral discretization and as Ra was increased, the number of Chebyshev modes used in the vertical discretization was increased from 33 to 513.   For each case, an error function was utilized as the initial condition for the diffusive concentration field
\begin{eqnarray}
C_i =  1 + \mbox{erf}\left(\dfrac{-(1-z)}{2\sqrt{D_it}}\right), \quad \mbox{for} \;\; 0 \le z < 1 \label{IC}
\end{eqnarray}
at time $t = 25/\mbox{Ra}^{2}$ or $t_{ad} = t\times\mbox{Ra}^2 = 25$ in advection-diffusion scaling \citep{Slim2014}, and a small random perturbation was added as a noise within the upper diffusive boundary layer to induce the convective instability. The solver has been verified in many previous investigations \citep{Wen2015, Wen2015thesis, Wen2015PRE, Shi2017, WenChini2018JFM}.

\section{Results}
\label{sec:Results}

\begin{figure}[ht]
    \centering
    \includegraphics[width=0.7\textwidth]{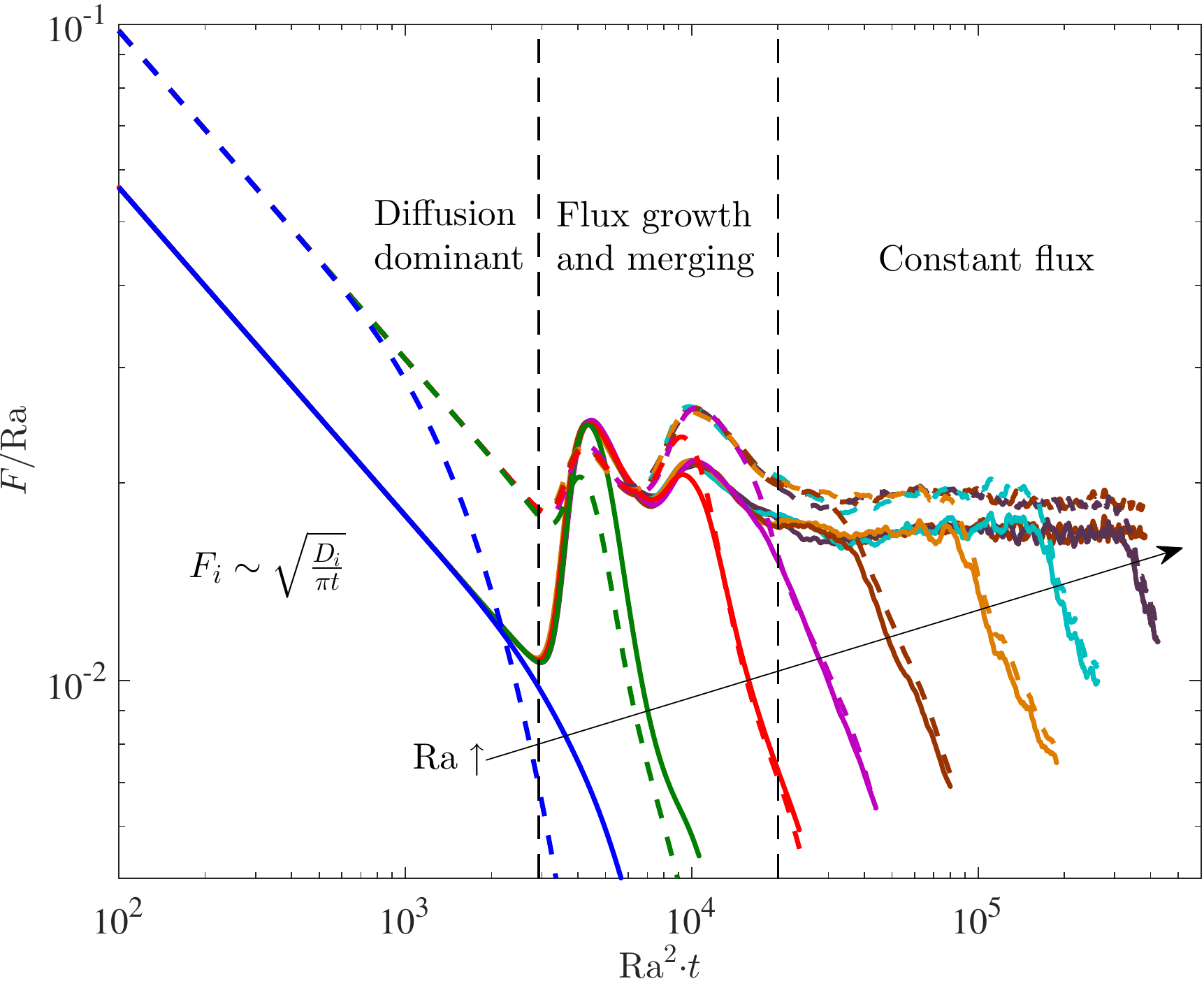} 
  \caption{Variation of the dissolution flux with time at $D_2 = 3$ for Ra $= 100$, 200, 500, 1000, 2000, 5000, $10^4$, $2 \times 10^4$ and $5 \times 10^4$.  Both the flux and time are rescaled following the advection-diffusion scaling to more evidently compare different regimes for different Ra.  The solid lines are for $C_1$ and dashed lines for $C_2$.  In the diffusion dominant regime, the flux for the solutes decays as $F_i \sim \sqrt{{D_i}/{(\pi t)}}$; for $2 \times 10^4 \lesssim t_{ad} \lesssim 16\text{Ra}$, the flow transitions to the constant-flux regime.}  \label{Flux_fixedD}
\end{figure}

Figure~\ref{Flux_fixedD} shows the variation of the dissolution flux with time for $D_2 = 3$ with increasing Ra.  Initially, the diffusion layer is far from the lower wall, the evolution of the purely diffusive concentration profile is universal (independent of Ra) in the advection-diffusion framework \citep{Slim2014} and follows Eq.~(\ref{IC}) so that $F_i \sim \sqrt{D_i/(\pi t)}$.  The top boundary layer becomes unstable when it is thick enough, thereby inducing convective fingers and making the flow deviate from the pure diffusion state \cite{Ennis-King2005, Xu2006, Riaz2006, Slim2010, Pau2010}.   As the nascent, independent-growing fingers penetrate the front of the diffusion layer, the plumes contact with more fresh water below the layer, leading to an increase of flux.  Subsequently, a secondary stability leads to lateral motions of the growing fingers and the flux growth regime ends when the neighboring fingers merge from the root.  After a series of plume mergers, which cause coarsening of the convective pattern, the flow transitions to a quasi-steady, constant-flux convective state with $F \sim \text{Ra}$, consistent with other high-Ra investigations of solutal convection \cite{Pau2010, Hidalgo2012, Slim2014} and thermal convection \cite{Doering1998, Otero2004, Hewitt2012, Wen2015, WenChini2018JFM} in porous media.  At the late time when the water is approximately $27\%$ saturated, the convection shuts down and the decay of the flux follows a simple box model \cite{Hewitt2013shutdown,Slim2013}.  In this study, we only focus on the dynamics quasi-steady constant-flux regime.

\begin{figure}[t]
    \centering
    \hspace{0.02in}\includegraphics[width=0.706\textwidth]{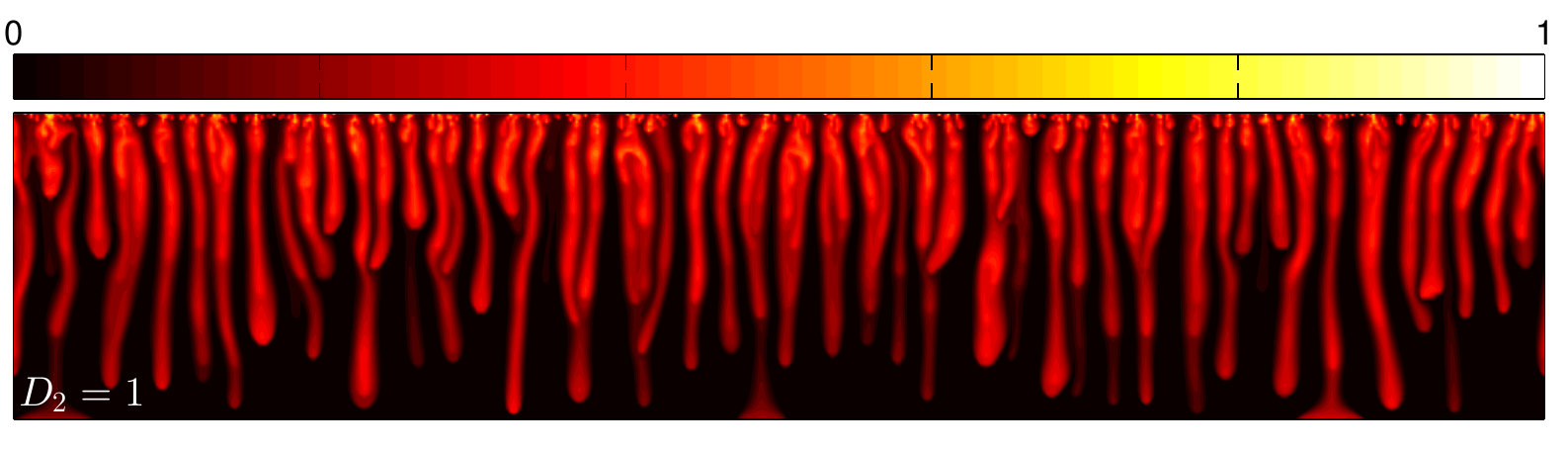}\\\vspace{-0.06in}
    \includegraphics[width=0.7\textwidth]{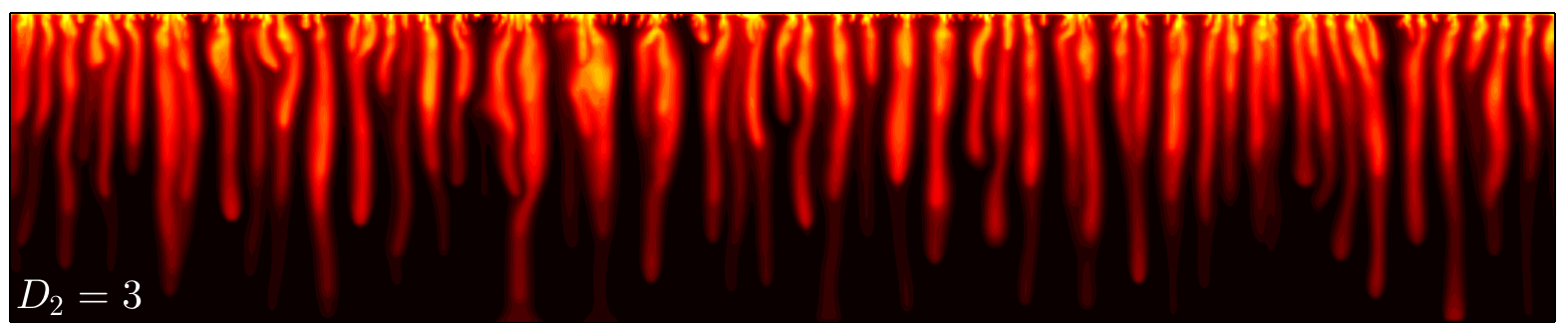}\\
    \includegraphics[width=0.7\textwidth]{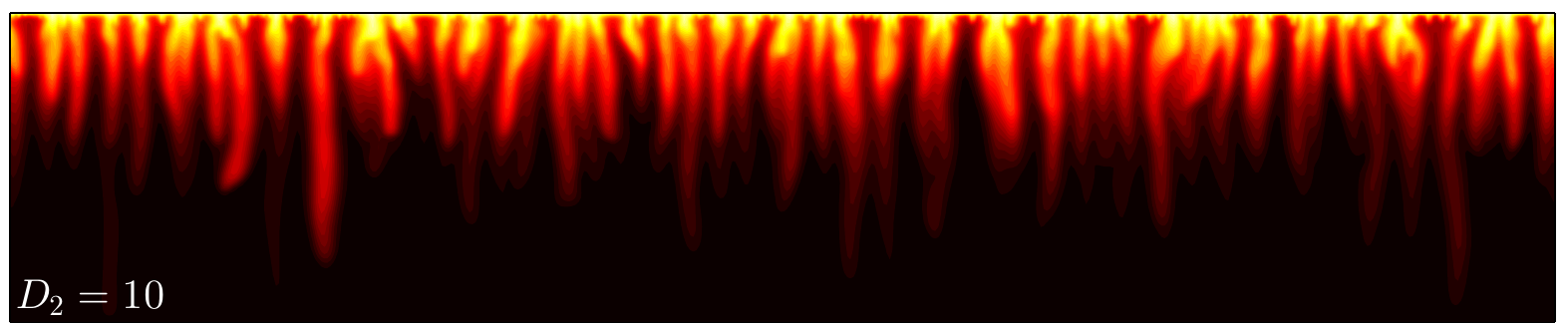}\\
    \includegraphics[width=0.7\textwidth]{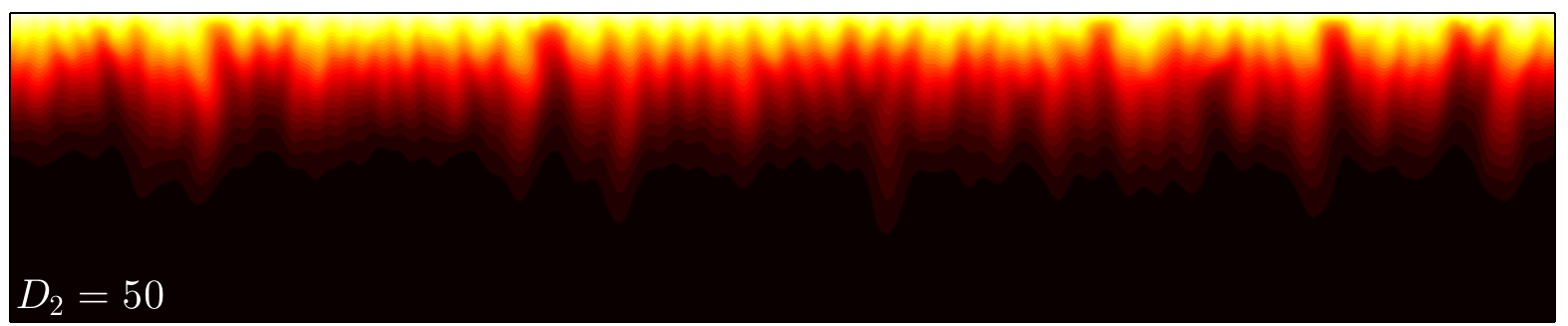}\\
    \includegraphics[width=0.7\textwidth]{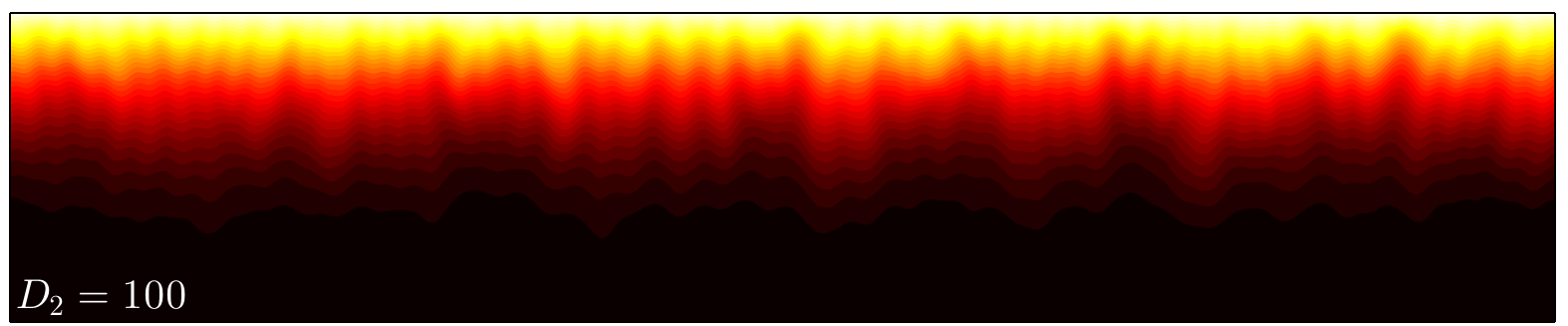}
    \caption{Concentration contours of $C_2$ at $t_{ad} = 8$Ra for different diffusivities at Ra $= 20000$.  For $D_2 > 1$, the downwelling plumes become much whiter, implying that more saturated solute is advected downward; moreover, as $D_2$ is increased,  the lateral concentration field is smoothed by diffusion and becomes nearly uniform for $D_2 \gtrsim 50$.}  \label{Cont_Ra20000_varD}
\end{figure}

\begin{figure}[t]
    \centering
    \includegraphics[width=0.7\textwidth]{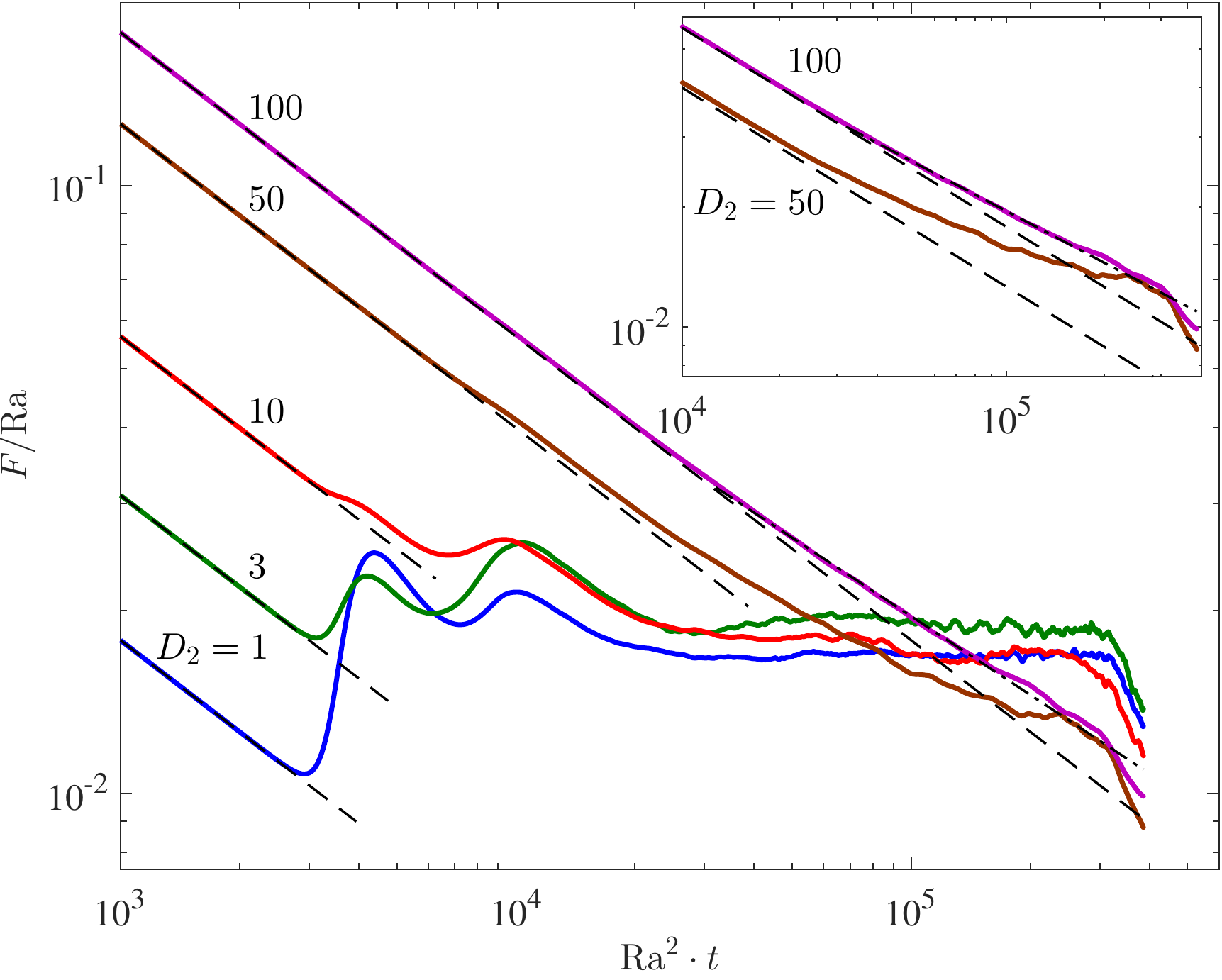}
  \caption{Variation of the dissolution flux with time for $C_2$ at Ra $= 20000$ for different diffusivities.  The dashed lines are for diffusion state and the inset shows a magnification of flux variation for $D_2 = 50$ and 100.  At large $D_2$, $C_2$ becomes horizontally averaged and just diffuses with a new effective diffusivity (see the dashed-dot line for $D_2 = 100$).}  \label{Fvst_varD}
\end{figure}

\begin{figure}[h!]
    \centering
    \includegraphics[width=0.7\textwidth]{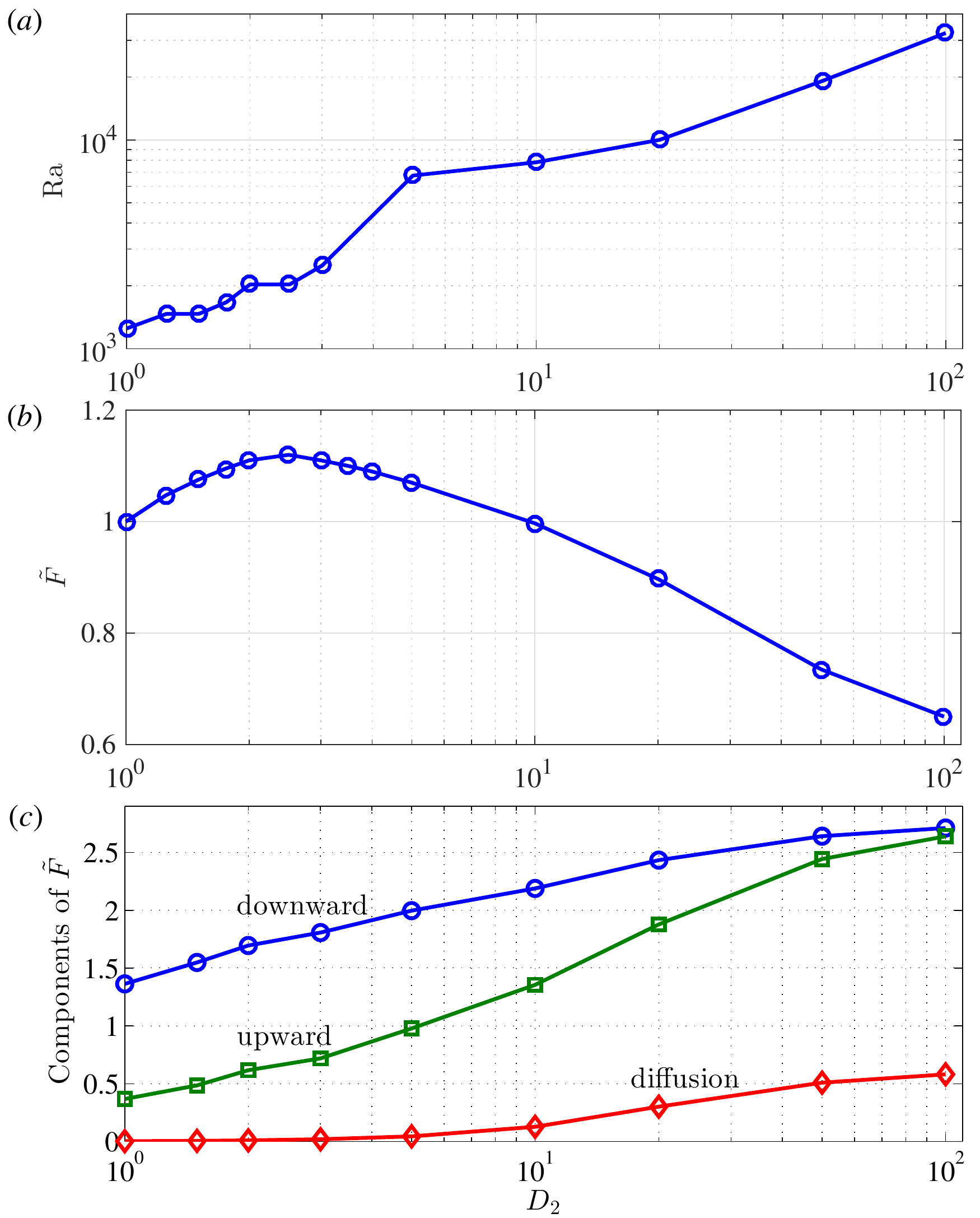}
  \caption{($a$): Approximated lower bound of Ra required to obtain the constant-flux regime from the simulations.  ($b$): Variation of the ratio of flux $\tilde{F}$ with $D_2$ in the constant-flux regime at sufficiently large Ra. ($c$):  Three components of $\tilde{F}$ through $z = 0.99$ at Ra = 50000. In ($a$), the existence of the constant-flux regime requires Ra $\sim O(10^3)$ for $D_2 < 5$.  In ($b$), through any horizontal plane, $\tilde{F} = $ $($downward advection $-$ upward advection $+$ diffusion$)$/$F_1$.}  \label{FratiovsDratio}
\end{figure}

As shown in Fig.~\ref{Flux_fixedD}, although $F_2$ generally follows the same trend with $F_1$ at $D_2 = 3$, they are not equivalent regardless of the magnitude of Ra.  For Ra $\lesssim 100$, diffusion dominates the dynamics, so $\tilde{F} = F_2/F_1 \sim \sqrt{D_2/D_1}$ before the diffusion front hits the bottom boundary.  Certainly, Rayleigh fractionation does not apply to the diffusion state.  Interestingly, even as Ra $\rightarrow \infty$, these two dissolution fluxes are still not equivalent, but the ratio $\tilde{F}$ converges to a constant value in the constant-flux regime at sufficiently large Ra.  Figure~\ref{Cont_Ra20000_varD} shows simulated concentration contours of $C_2$ for different $D_2$ at Ra $= 20000$.  In this case, the concentration contours of $C_2$ basically retain the finger features for $D_2 < 5$.  However, the increasing diffusivity gradually smooths the long and thin fingers and at sufficiently large $D_2$, makes the concentration field almost uniformly distributed in $x$ and just diffuse with a new scaling $F_2 \sim t^{-\gamma}$ with $0 < \gamma < 1/2$ (see $D_2 = 100$ in Fig.~\ref{Fvst_varD}).  As also shown in Fig.~\ref{Fvst_varD}, for fixed large Ra and at small $D_2$, $F_2$ generally follows the same variation of $F_1$.  Nevertheless, the increasing $D_2$ will postpone the occurrence of the constant-flux regime (see $D_2 = 10$), implying that a larger $D_2$ requires corresponding larger Ra's to obtain the constant-flux regime before the convection shuts down (see Fig.~\ref{FratiovsDratio}$a$). 

 As discussed above, for each fixed $D_2$, the finger features and constant-flux regime can be retained at sufficiently large Ra.  Figure~\ref{FratiovsDratio}($b$) shows the ratio of fluxes between tracer and CO$_2$ in the constant-flux regime as a function of $D_2$.  At $D_2 = 1$, the two solutes are equivalently transported so that $\tilde{F} \equiv 1$; interestingly, for $D_2 \le 2.5$, the increase of $D_2$ enhances the convective mixing of the solute $C_2$, e.g. the flux $F_2$ is nearly $12\%$ increased at $D_2 = 2.5$; for $D_2 > 2.5$, however, $\tilde{F}$ decreases as $D_2$ is increased, and for $D_2 > 10$, $\tilde{F} < 1$, implying that the large diffusivity reduces the mixing efficiency of $C_2$.  Since the flow field is only set by $C_1$, the increase of diffusivity thickens the top diffusion boundary layer (see Fig.~\ref{Cont_Ra20000_varD}), so that more saturated brine is advected downward by fingers from the upper layer.  Therefore, moderate increase of the diffusivity could increase the dissolution rate of the tracer.  Nevertheless, due to the conservation of mass, relatively fresh brine rises to the top through the upwelling flows. As $D_2$ is increased, the strong lateral diffusion smooths the high concentrations to the sides, leads to a leakage from the downwellings into the upwellings (see Fig.~\ref{FratiovsDratio}$c$), and thereby significantly decreases the dissolution rate.
 

At large Rayleigh number, the solutal convection in the porous layer appears in the form of narrow fingers with the wavelength $L_m$ shrinking as a power-law scale of Ra; namely, the mass transport is generally performed through these downwelling and upwelling plumes.  To a certain extent, this phenomenon is analogous to a Taylor (or Taylor--Aris) dispersion problem \cite{Taylor1953, Aris1956}, where spread of the solute in a 2D channel is enhanced by the axial flow.  In the CO$_2$-tracer `dispersion' problem, the channel has a height 1 and width $L_m$. Away from the top and bottom boundary layers, the horizontal velocity $u$ is negligible and the vertical (axial) velocity can be approximated using $w = {W_0}\cos(kx)$, where $W_0 = a$Ra with the constant pre-factor $a$ and $k = 2\pi/L_m$ is the fundamental wavenumber.  As the tracer is advected downward, it also diffuses to both sides and the amplitude of the concentration fluctuation (i.e., deviations from the horizontal mean) decays as the exponential rate $e^{-D_2k^2t}$, so that the time required by diffusion to well smooth the fluctuation term (down to $1\%$) over $L_m$  is $t_1 = {2\ln10}/{(D_2k^2)}$.  Moreover, the study by Slim~\citep{Slim2014} indicates that the fingertips travel with a constant speed $0.13$Ra before hitting the lower boundary.  Therefore, the time required for $C_2$ to be advected downward across the \emph{same} length $L_m$ is $t_2 = 2\pi/(0.13Rak)$.  Hence, to obtain a horizontally uniform concentration field, it requires at least $t_1 \le t_2$, i.e. $D_2 \ge \frac{0.13\ln10}{\pi}{\text{Ra}}/{k} = \frac{0.13\ln10}{2\pi^2}\text{Ra}L_m$, or $D_2 \sim O(\text{Ra}L_m)$.  For instance, at Ra $= 20000$, $L_m \lesssim 0.14$ before the shut-down regime, so  $D_2 \ge 42.5$, quantitatively consistent with the results shown in Fig.~\ref{Cont_Ra20000_varD}.  It will be shown below $D_2 \sim O(\text{Ra}L_m)$ actually corresponds to $O(1)$ P\'{e}clect number in the dispersion model. 

Renormalize the variables $\tilde{t}$ = Ra$t$, $\tilde{w}$ = $w/{W_0}$,  $X = x/\varepsilon$, where $\varepsilon = L_m \sim \text{Ra}^{-0.4}$ is a small parameter at large Ra \citep{Hewitt2012}, so that the time and velocity fields are transformed from diffusion scales to convection scales.  Finally, Eq.~(\ref{Solute_nondim}) for $C_2$ becomes 
\begin{eqnarray}
	\text{Pe}\;\varepsilon\left(\dfrac{1}{a}\dfrac{\partial C_2}{\partial \tilde{t}} + \tilde{w}\dfrac{\partial C_2}{\partial z}\right) = \left(\dfrac{\partial^2}{\partial X^2} + \varepsilon^2 \dfrac{\partial^2}{\partial z^2} \right)C_2, \label{Model_C2_2}
\end{eqnarray}
where the constant value $a$ and $\tilde{w} = \cos(2\pi X)$ are of order unity, and the P\'{e}clet number
\begin{eqnarray}
\text{Pe} = \dfrac{a\text{Ra}L_m}{D_2} = \dfrac{W_0\varepsilon}{D_2} \equiv \dfrac{\varepsilon^2/D_2}{\varepsilon/W_0} \label{Peclet}
\end{eqnarray}
denotes the ratio between the advective and diffusive (dispersive) time scales.  From our previous analysis, the horizontally uniform concentration requires $D_2 \sim O(\text{Ra}L_m)$, namely, Pe $\sim O(1)$.  For any $D_2 \sim o(\text{Ra}L_m)$, e.g. $D_2 \sim O(1)$, Pe $\rightarrow \infty$ as Ra $\rightarrow \infty$, and then the concentration field appears in the form of apparent fingers at sufficiently large Ra.  

\section{Conclusions}
\label{sec:Conclusions}

The fundamental role of diffusion in mass or heat transport has been studied extensively in the convection problem.  In the `ultimate' high-Ra regime, the analysis based on the assumption that the molecular diffusive transport is negligible when Ra = advection/diffusion $\gg 1$ \cite{Spiegel1971, Grossmann2000} generally yields an invalid asymptotic $F$--Ra scaling \cite{Whitehead2011}.  For the CO$_2$-tracer, solutal convection problem, our study indicates that the mass transport also \emph{depends} on the molecular diffusion, which is in contradiction to the classical Rayleigh fractionation assumption that the fractionation of different components is only determined by their solubility.  When the solubility constants of the two components are close, i.e. $K_1/K_2 \sim O(1)$, the difference between $F_1$ and $F_2$ might have a first-order effect on the fractionation. However, for the noble gases He, Ne, and Ar which are usually used as tracers to identify CO$_2$ dissolution in carbon sequestration, the ratio of the solubility constant $K_1/K_2 > 20$, so that the $O(1)$ variation of $F_1/F_2$ will not affect the approximation $\mathcal{F} \approx 1 - r/r^0$ and the Rayleigh fractionation is realized.

\section*{Acknowledgement} 
This work was supported as part of the Center for Frontiers of Subsurface Energy Security, an Energy Frontier Research Center
funded by the U.S. Department of Energy, Office of Science, Basic Energy Sciences under Award \# DE-SC0001114. B.W. acknowledges the
Peter O'€™Donnell, Jr. Postdoctoral Fellowship in Computational Engineering and Sciences at the University of Texas at Austin.

\appendix
 \section{Variation of gas composition}
 The change of $i$-th component in the gas field can be expressed as 
\begin{eqnarray}
 \dfrac{dn_{i,g}}{dt} = -qF_iC_{is}, \label{ODE_Rayleigh}
\end{eqnarray}
where $q=D^*_1A/H$.  For multicomponent ideal gas,
\begin{eqnarray}
	P_{i,g}V_g = n_{i,g}RT \;\; \Rightarrow \;\; P_gV_g = (\sum n_{i,g})RT, \label{ideal_gas}
\end{eqnarray}
where $P_{i,g}$ is the partial pressure of the $i$-th component, $V_g$ is the total gas volume,  $R$ is the universal gas constant, $T$ is the absolute temperature, and $P_g=\sum P_{i,g}$ is the total gas pressure. From Henry's law,
\begin{eqnarray}
 P_{i,g} = \frac{C_{is}}{K_{i}}. \label{Henry_law}
\end{eqnarray}
The equations~(\ref{ideal_gas}) and (\ref{Henry_law}) yield
\begin{eqnarray}
 C_{is} = K_{i}P_{i,g} = \frac{K_{i}RT}{V_g}n_{i,g} = K_{i}P_g\frac{n_{i,g}}{n_{1,g} + n_{2,g}}. \label{Henry_law2}
\end{eqnarray}
Substituting (\ref{Henry_law2}) into (\ref{ODE_Rayleigh}) gives 
\begin{eqnarray}
 \dfrac{dn_{i,g}}{dt} = -qF_i\frac{K_{i}RT}{V_g}n_{i,g} = -qF_iK_iP_g\frac{n_{i,g}}{n_{1,g} + n_{2,g}}. \label{ODE_Rayleigh2}
\end{eqnarray}
Then, we have 
\begin{eqnarray}
 \dfrac{dn_{1,g}}{dn_{2,g}} = \alpha\frac{n_{1,g}}{n_{2,g}}, \label{ODE_Rayleigh3}
\end{eqnarray}
where $\alpha = F_1K_1/(F_2K_2)$.  For a quasi-steady convective system, the dissolution flux $F_i$ is fixed, so that $\alpha$ is constant. Then
\begin{eqnarray}
 \ln\dfrac{n_{1,g}}{n_{1,g}^0} = \alpha \ln\dfrac{n_{2,g}}{n_{2,g}^0} \quad \mbox{or} \quad \dfrac{n_{1,g}}{n_{1,g}^0} =  \left(\dfrac{n_{2,g}}{n_{2,g}^0}\right)^\alpha. \label{ODE_Rayleigh4}
\end{eqnarray}
Namely, 
\begin{align}
r\equiv \dfrac{n_{1,g}}{n_{2,g}} =  \dfrac{n_{1,g}^0 \cdot n_{2,g}^{\alpha - 1}}{(n_{2,g}^0)^\alpha}  \Rightarrow n_{2,g} = \left(\dfrac{{n_{2,g}^0}^\alpha \cdot r}{n_{1,g}^0}\right)^{\frac{1}{\alpha-1}}. \label{ODE_Rayleigh5}
\end{align}
Therefore, the fraction of dissolved CO$_2$ into water is
\begin{eqnarray}
\mathcal{F} \equiv 1 - \dfrac{n_{1,g}}{n_{1,g}^0} = 1 - \dfrac{n_{1,g}}{n_{2,g}}\cdot\dfrac{n_{2,g}}{n_{1,g}^0} = 1 - (r/r^0)^{\frac{\alpha}{\alpha - 1}}. \label{fraction_CO2}
\end{eqnarray}
Actually, (\ref{fraction_CO2}) is a generic form which is valid for both constant $P_g$ and constant $V_g$.  When $\alpha \gg 1$,
\begin{eqnarray}
\mathcal{F} \approx 1 - (r/r^0). \label{fraction_CO2_2}
\end{eqnarray}

\bibliographystyle{model1-num-names}
\bibliography{WHref}

\end{document}